\documentclass[a4paper,11pt]{article}
\usepackage{pos}
\usepackage{xspace}
\usepackage{siunitx}
\usepackage{hyperref}
\usepackage{ulem}

\newcommand{\plenum}{PLE$\nu$M\xspace}
\newcommand{\txs}{TXS~0506+056\xspace}
\newcommand{\ngc}{NGC~1068\xspace}
\newcommand{\drv}[1]{\ensuremath{{\rm d}#1}}
\newcommand{\aeff}{\ensuremath{A_{\rm eff}}}
\newcommand{\ecut}{\ensuremath{E_{\rm cut}}}
\newcommand{\const}{\ensuremath{\text{const.}}}

\DeclareSIUnit{\EeV}{EeV}
\DeclareSIUnit{\PeV}{PeV}
\DeclareSIUnit{\TeV}{TeV}
\DeclareSIUnit{\Mpc}{Mpc}
\DeclareSIUnit{\Gpc}{Gpc}
\DeclareSIUnit{\erg}{erg}
\DeclareSIUnit{\year}{yr}

\title{\plenum: A global and distributed monitoring system of high-energy astrophysical neutrinos}
 \ShortTitle{\plenum}

\author*[a]{Lisa Schumacher}
\author[a]{Matthias Huber}
\author[b]{Matteo Agostini}
\author[c]{Mauricio Bustamante}
\author[d]{Foteini Oikonomou}
\author[a]{Elisa Resconi}

\affiliation[a]{Technical University Munich, Garching, GER}

\affiliation[b]{University College London, London, UK}

\affiliation[c]{Niels Bohr Institute, Copenhagen, DEN}

\affiliation[d]{Norwegian University of Science and Technology, Trondheim, NOR}

\emailAdd{lj.schumacher@tum.de}

\abstract{High-energy astrophysical neutrinos, discovered by IceCube, are now regularly observed, albeit at a low rate due to their low flux. As a result, open questions about high-energy neutrino astrophysics and particle physics remain limited by statistics at best, or unanswered at worst. Fortunately, this situation will improve soon: in the next few years, a host of new neutrino telescopes, currently under planning and construction, will come online. It is natural to combine their collected observing power: we propose the Planetary Neutrino Monitoring System (\plenum), a concept for a global repository of high-energy neutrino observations, in order to finally give firm answers to open questions. \plenum will reach up to four times the exposure available today by combining the exposures of current and future neutrino telescopes distributed around the world -- IceCube, IceCube-Gen2, Baikal-GVD, KM3NeT, and P-ONE. Depending on the declination and spectral index, \plenum will improve the sensitivity to astrophysical neutrinos by up to two orders of magnitude. We present first estimates on the capability of \plenum to discover Galactic and extragalactic sources of astrophysical neutrinos and to characterize the diffuse flux of high-energy neutrinos in unprecedented detail.}

\FullConference{37$^{\rm{th}}$ International Cosmic Ray Conference (ICRC 2021)\\
		July 12th -- 23rd, 2021\\
		Online -- Berlin, Germany}

\begin{document}
\maketitle

\section{Introduction}
The discovery of high-energy astrophysical neutrinos made by IceCube marked the beginning of high-energy neutrino astronomy.
By now, IceCube has gathered more than 10 years' worth of neutrino data and found two neutrino-source candidates: the blazar \txs \cite{icecubePriorTXS2018,icecubeMultiMessengerTXS2018} and the Seyfert Galaxy \ngc \cite{tenyrIC_PS:2019}.  
But key questions about the origin of the high-energy astrophysical neutrinos~\cite{Stettner:2019TL} remain unanswered  due to statistical uncertainties.  
Without a significant increase in the detection rate of astrophysical neutrinos, progress in answering these questions might be slow.

Figure~\ref{fig:aeff_ic_pl} shows the effective area $A_{\rm eff}$ of IceCube for through-going muon tracks initiated by muon neutrinos, which illustrates why the event rate is presently limited:
For tracks coming from the Southern Hemisphere ($\sin \delta < 0$), $A_{\rm eff}$ is suppressed due to the need to suppress the overwhelming rate of atmospheric muons.  
For tracks coming from the Northern Hemisphere ($\sin \delta > 0$), $A_{\rm eff}$ is suppressed at high energies because the Earth becomes increasingly opaque for neutrinos with energies above $\sim$\SI{0.1}{\PeV}.
Thus, even though IceCube sees the whole sky in neutrinos, its effective area for high-energy neutrinos depends strongly on declination: it is maximal close to the horizon and up to about \SI{30}{\degree} North in declination.
Therefore, not only is the full-sky neutrino rate suppressed, but, also, IceCube is not well suited to observe sources in the Southern Hemisphere, where, e.g., the Galactic Center (GC) is located.

Fortunately, a natural solution is within reach:
In the next decades, multiple neutrino telescopes distributed across the globe will come online
-- IceCube-Gen2~\cite{Aartsen_2021:gen2}, KM3Net~\cite{km3netIntent2016}, P-ONE~\cite{agostiniPacificOceanNeutrino2020}
and Baikal-GVD~\cite{baikalGVD2019} --
greatly increasing the rate of neutrino detection, especially in the Southern Hemisphere.
We propose the Planetary Neutrino Monitoring System (\plenum), a concept for a global repository of high-energy neutrino observations made by current and future neutrino telescopes.
In the following sections, we present first estimates for the improvements with \plenum regarding the search for individual neutrino sources and the characterization of the diffuse astrophysical neutrino flux.
The programming code for this study is under development and available at 
\href{https://github.com/mhuber89/Plenum}{https://github.com/mhuber89/Plenum}.

\section{\texorpdfstring{\plenum}{PLEnuM}: A Planetary Neutrino Monitoring System}

In a neutrino telescope, the expected number of detected neutrinos as a function of the livetime, $T_{\rm live}$, and solid angle $\Delta \Omega$, can be calculated from the effective area, \aeff,  and the neutrino flux, $\frac{\drv \Phi}{\drv E}$, via
\begin{equation}
\label{eq:n}
N_{\nu} = T_{\rm live} \cdot \int_{\Delta \Omega}\drv \Omega \, \drv E \, \int_{E_{\min}}^{E_{\max}} \drv E \, \aeff \left( E, \sin(\delta)\right) \cdot \frac{\drv \Phi}{\drv E},
\end{equation}
where $E$ is the neutrino energy.
The total number of neutrinos detected with multiple detectors is the sum of the number detected by each detector.
This is equivalent to replacing the single-telescope effective area in Eq.~\eqref{eq:n} by the sum of the individual effective areas.
Therefore, we study the potential of \plenum based on combining the effective areas of (hypothetical) telescopes at different locations on Earth.

We use the effective area for muon neutrinos of the 10-year data set published by IceCube as a basis \cite{tenyrIC_PS:2019, ICdataRelease2021web}, which is optimized for point-source searches due to the good angular resolution of below $1^\circ$.
Rather than using the estimated effective areas published by the planned experiments,
we employ a simplified strategy to estimate the effective area of \plenum:
we assume that at each location of the four constituting detectors -- IceCube, KM3Net~\cite{km3netIntent2016}, P-ONE~\cite{agostiniPacificOceanNeutrino2020}
and Baikal-GVD~\cite{baikalGVD2019} -- there is a detector with IceCube's effective area.
In order to model a contribution inspired by IceCube-Gen2~\cite{Aartsen_2021:gen2}, we assume a detector at the South Pole with an effective area 7.5 times larger than that of IceCube\footnote{This is motivated by the claim that the discovery potential for point-like neutrino sources is a factor of 5 better than that of IceCube~\cite{Aartsen_2021:gen2}. Together with Eq.~\eqref{eq:livetime_scaling} this yields the factor 7.5 in \aeff.}.
Based on these simplifications, all calculations presented here are estimates to be refined in future works.

The effective areas in equatorial coordinates depend on time, in case the detector is not located at the Geographic South Pole.
In order to obtain the effective areas independent of time, we integrate the local effective areas over one full sidereal day. 
Note that this integration is only a valid approach when we study time-integrated properties of \plenum. 
The effective area of \plenum is then the sum of the constituting integrated effective areas.
If different detectors have different livetimes, we use Eq.~\eqref{eq:n} to scale each detector contribution for calculating the total number of detected neutrinos.

\begin{figure}
\centering
\includegraphics[width=\textwidth]{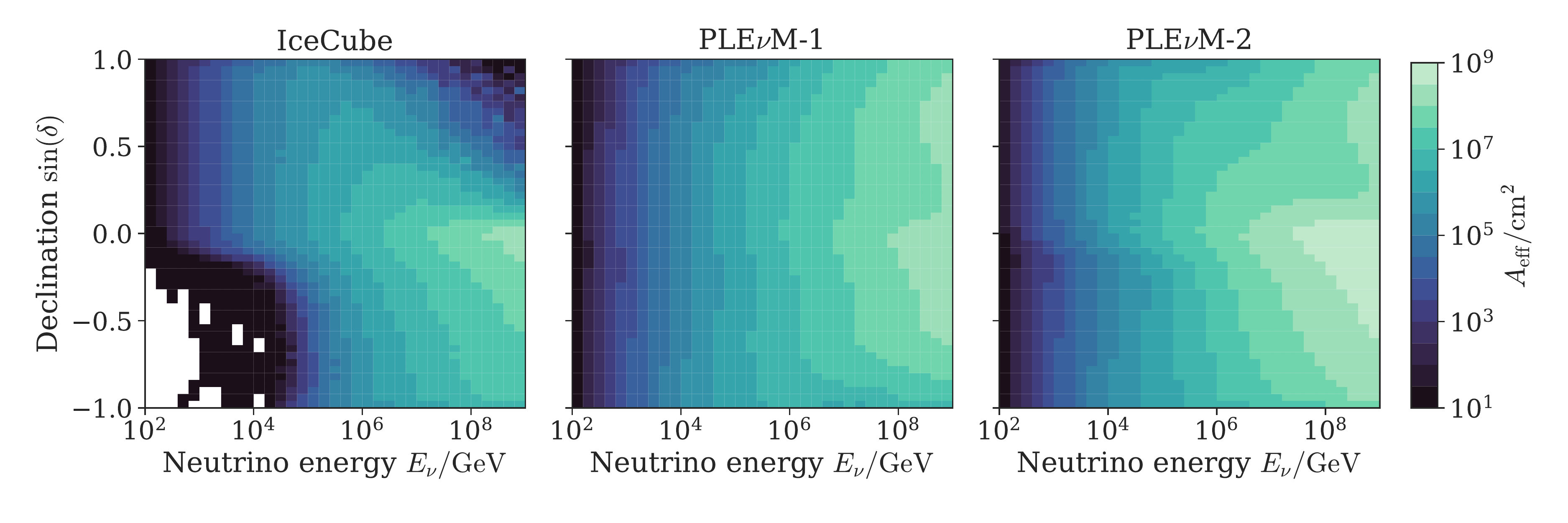}
\caption{Comparison of the effective area for through-going muon tracks of IceCube, \plenum-1 and \plenum-2 as a function of energy and sine of declination.
\plenum-1 consists IceCube and IceCube-like detectors at the locations of P-ONE, KM3NeT and Baikal-GVD. 
For \plenum-2, we assume a detector that is 7.5 times larger than IceCube instead of IceCube's contribution to \plenum.}
\label{fig:aeff_ic_pl}
\end{figure}

We study two different configurations of \plenum: one consisting of IceCube and IceCube-like detectors at the locations of KM3NeT, P-ONE and Baikal-GVD (\plenum-1), and another one with a contribution of a hypothetical detector that has a 7.5 times larger effective area instead of IceCube (\plenum-2).
Figure~\ref{fig:aeff_ic_pl} shows the effective area of IceCube, \plenum-1 and \plenum-2: 
In contrast to the strong declination dependence for IceCube, the effective areas of \plenum-1 and \plenum-2 cover the whole sky more evenly. 

\begin{figure}
\centering
\includegraphics[width=\textwidth]{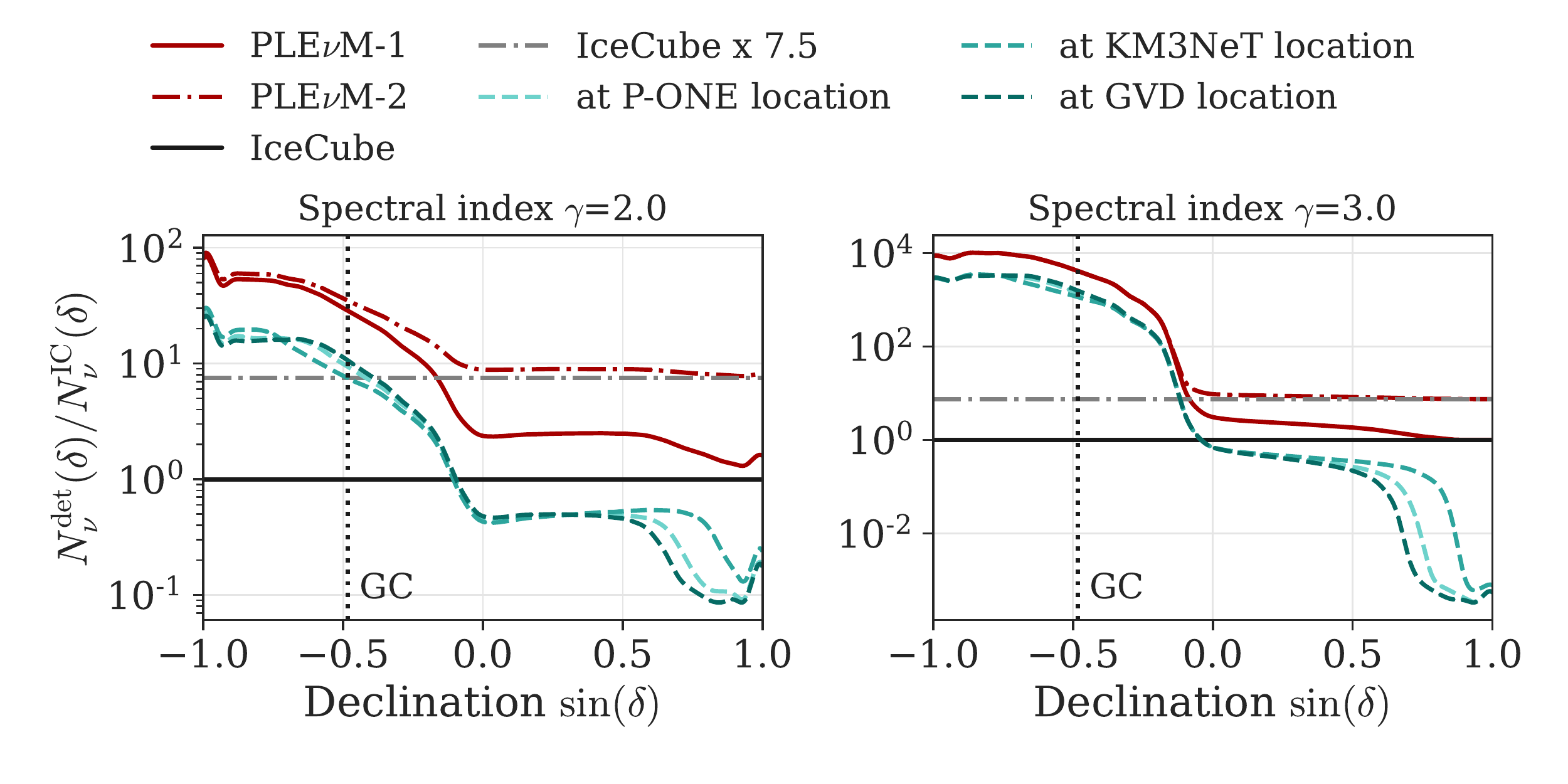}
\caption{Number of neutrinos detected by all constituent detectors and \plenum-1/2 for through-going muon tracks relative to neutrinos detected by IceCube at the respective sine of declination. The two plots show the ratio calculated based on spectral indices of $\gamma=2.0$ and $\gamma=3.0$ of the astrophysical neutrinos. The declination of the Galactic Center is marked with a vertical, dotted line.}
\label{fig:improv_det_eff}
\end{figure}

In order to quantify the improvement in neutrino detection with \plenum, we look at the ratio of the number of neutrinos detected by a particular detector or \plenum to the number of neutrinos detected by IceCube at the same declination, i.e., $N_{\nu}^{\rm det}(\delta) / N_{\nu}^{\rm IC}(\delta)$.
Figure~\ref{fig:improv_det_eff} shows this ratio
for different values of the spectral index. 
At the declination of the Galactic Center (GC) in the Southern Hemisphere, \plenum-1 and \plenum-2 can improve the detection efficiency by a factor of about 30 for $\gamma=2.0$ and by more than three orders of magnitude for $\gamma=3.0$.


\subsection{Point sources}

\begin{figure}
\includegraphics[width=\textwidth]{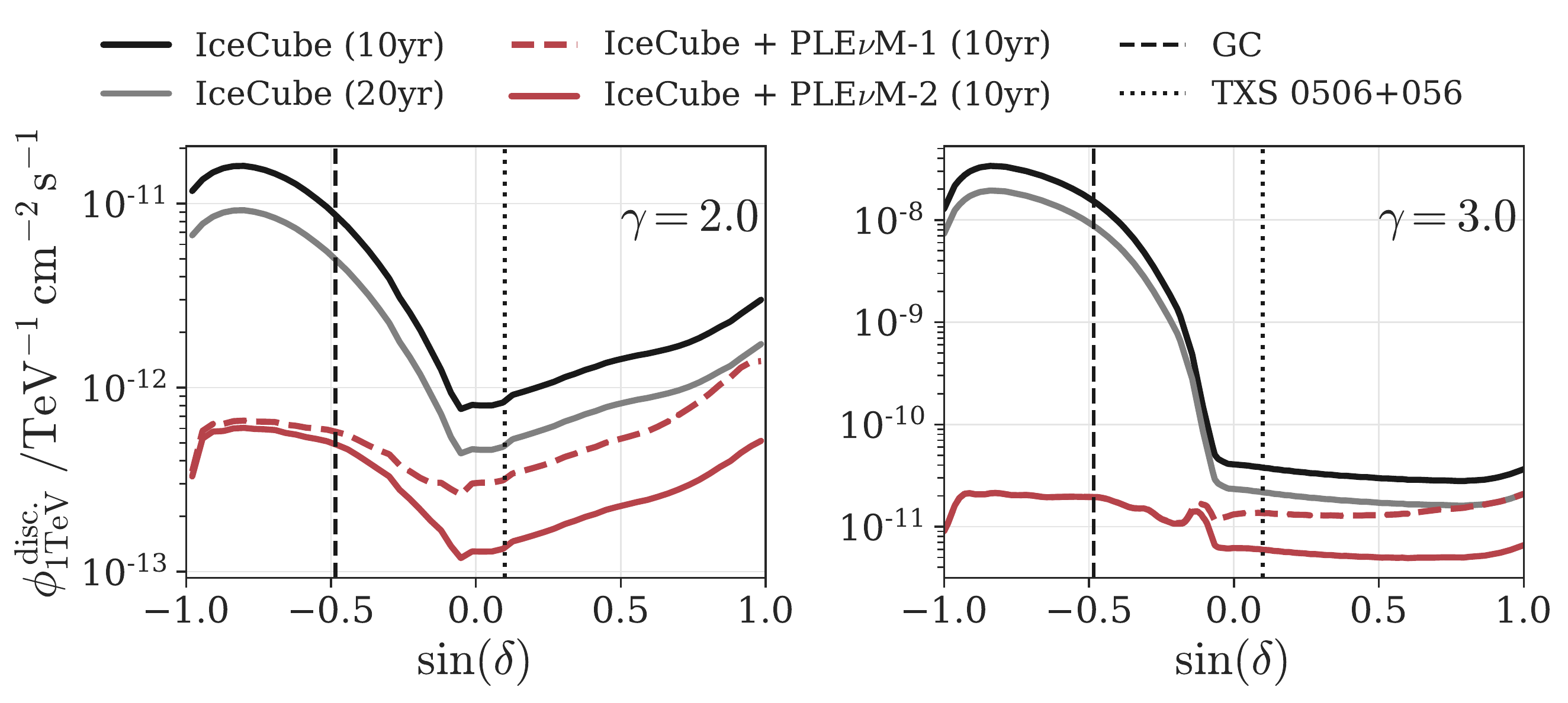}
\caption{Comparison of discovery potentials (DP) for two spectral indices: $\gamma=2.0$ (left) and $\gamma=3.0$ (right).
The DP is calculated for an neutrino flux per source with a spectrum of $\drv \Phi / \drv E = \Phi^{\mathrm{disc.}} \cdot (E/\SI{1}{\TeV})^{-\gamma}$ at \SI{1}{\TeV}.
Shown are the DPs based on the 10yr PS analysis by IceCube~\cite{tenyrIC_PS:2019} (black), an estimate with a livetime of 20 yrs (gray), \plenum-1 including IceCube (dashed red) and \plenum-2 including IceCube $\times 7.5$ (solid red).}
\label{fig:improv_dp}
\end{figure}

The $5\sigma$ discovery potential (DP) is used in point-source analyses in IceCube to quantify the performance of the analysis and quality of the data set.
The DP is the neutrino flux per source with a spectrum of $\drv \Phi / \drv E = \Phi^{\mathrm{disc.}} \cdot (E/\SI{1}{\TeV})^{-\gamma}$ at \SI{1}{TeV} that is needed to claim a $5\sigma$ discovery with respect to a uniform distribution of neutrinos.
Motivated by Fig.~15 in~Ref.~\cite{aartsen7yrIntegratedPS2017}, we assume that the DP in IceCube improves with the livetime of the data set approximately with
\begin{equation}
\label{eq:livetime_scaling}
    \frac{\phi^{\mathrm{disc.}}_{\rm IC} (T_{\rm live}=\SI{10}{\year})} {\phi^{\mathrm{disc.}_1} (T_{\rm live}=T_1)} =
    \begin{cases} 
    \left( \frac{T_0}{T_1} \right) ^{-0.8}  & \text{if } \aeff=\const\\
    \left( \frac{A_{\rm eff, 0}}{T_{\rm eff, 1}} \right) ^{-0.8}  & \text{if } T_{\rm live}=\const
    \end{cases} \\
\end{equation}
Based on Eq.~\eqref{eq:n}, we derive that this relation can also be translated to a scaling of effective areas, or a combination of scaling the livetime and effective area.
We use the results and the public data set presented in Ref.~\cite{tenyrIC_PS:2019}
together with Eq.~\eqref{eq:livetime_scaling} to
estimate the improvement in DP with \plenum.
Figure~\ref{fig:improv_dp} shows the DP of 10 years of IceCube data presented in Ref.~\cite{tenyrIC_PS:2019}, as well as estimates for the DP using 20 years of IceCube data, 
and a combination of IceCube and \plenum data with 10 years of livetime each.
At the declination of the Galactic Center, the DP improves by one order of magnitude for a spectral index of $\gamma=2.0$, and by almost three orders of magnitude for a spectral index of $\gamma=3.0$.
Around the horizon and in the Northern Hemisphere, our estimate for IceCube $\times \, 7.5$ contributes significantly to the improvement in DP.

\subsection{Diffuse flux}
Besides the search for individual sources, the diffuse spectrum of astrophysical neutrinos
can shed light on the features and populations of neutrino sources.
As a case study, we compare two hypotheses for the neutrino spectrum:
a single power law, as often assumed in IceCube analyses, and a power law with an exponential cutoff in energy.
The \plenum effective area used for the diffuse flux analyses is based on the IceCube effective area which is set to zero below $-5^\circ$ in declination.
This is done in order to mimic the cut in declination applied in the event selection of the standard diffuse analysis of through-going muons~\cite{Stettner:2019TL},
which requires a high-purity sample by strongly suppressing muon contamination from the Southern Hemisphere\footnote{The effective areas and event selections of Ref.~\cite{Stettner:2019TL} and~\cite{tenyrIC_PS:2019} are similar, but not identical.}.

We perform a binned maximum likelihood analysis based on the function
\begin{equation}
\label{eq:diff_likelihood}
    \mathcal{L}({\rm data}~k~ |~{\rm hypothesis}~\mu) 
    = \prod_{{\rm bin\,}ij}^{N_{\rm bins}} \frac{\mu_{ij}^{k_{ij}}}{k_{ij}!}\cdot
    \exp \left( -\mu_{ij} \right), 
\end{equation}
where $\mu$ is the neutrino expectation and $k$ is the observed data.
The indices $ij$ run over the reconstructed energy and declination bins, respectively.
The expected number of events is a sum of expected atmospheric neutrinos and astrophysical neutrinos given a the specific hypothesis. 
The expectations in bins of reconstructed energy are obtained by applying a smearing matrix provided in IceCube's data release~\cite{ICdataRelease2021web}, while we neglect the uncertainty in arrival direction which is smaller than the bin size in declination.
We estimate the discovery power by using a likelihood ratio test and its asymptotic $\chi^2$ behavior~\citep{wilks1938} in combination with representative Asimov data sets~\cite{Cowan_2011}

\begin{figure}
    \centering
    \includegraphics[width=0.49\textwidth]{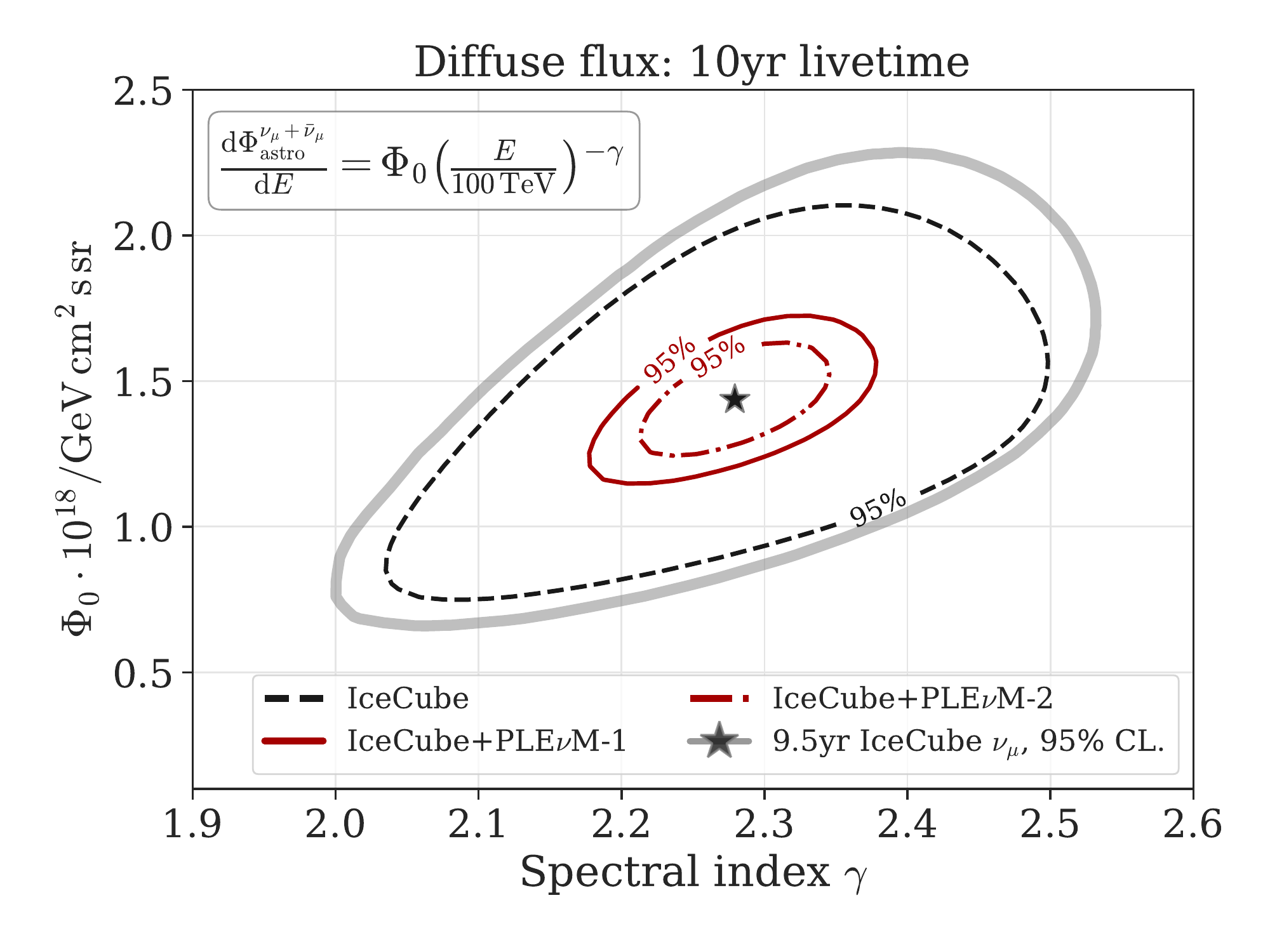}
    \includegraphics[width=0.49\textwidth]{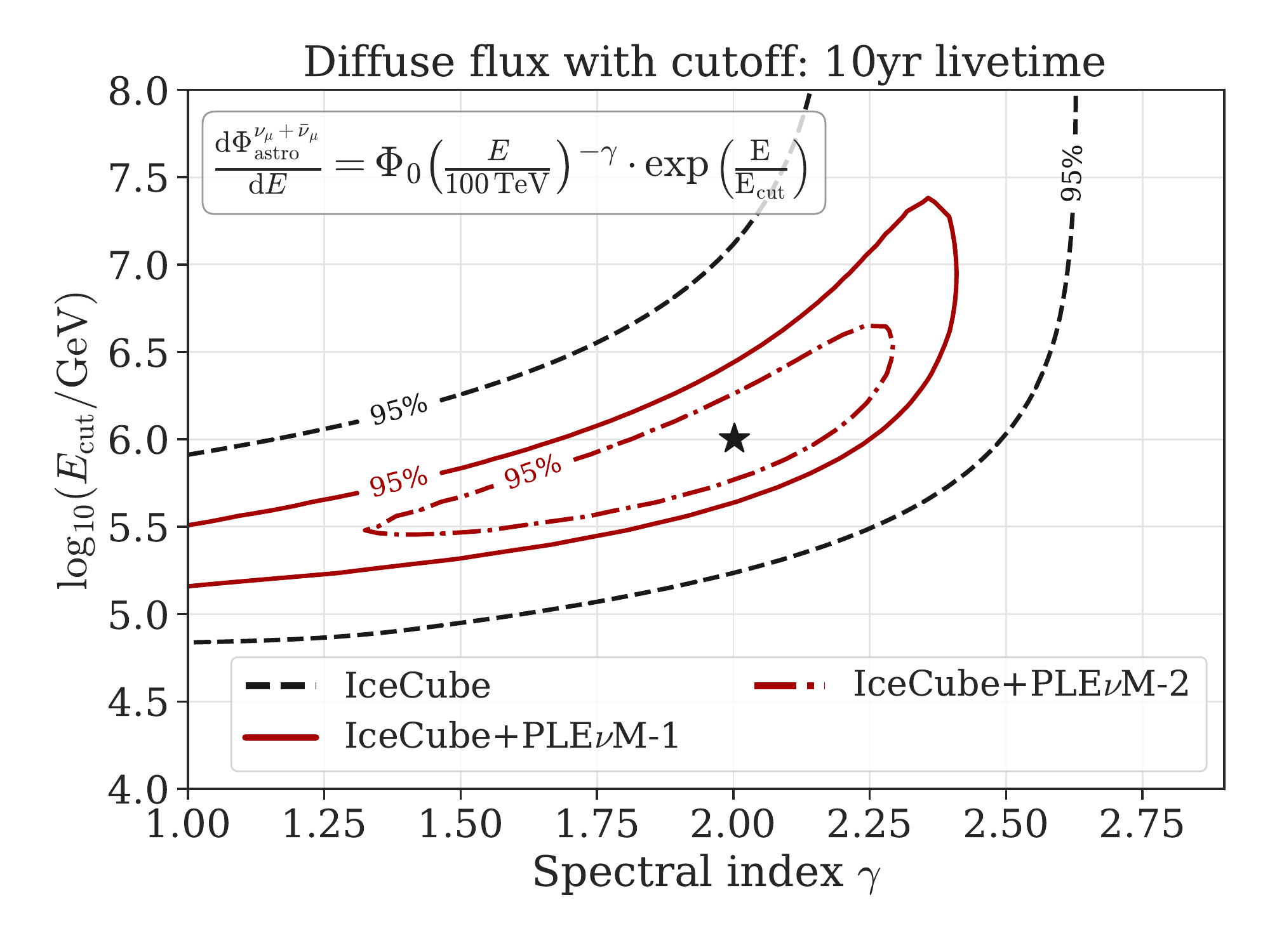}
    \caption{Left: Confidence intervals obtained from the diffuse astrophysical fit on Asimov data with a power-law spectrum for IceCube, \plenum, and, as a cross check, the official IceCube analysis on a similar data set \cite{Stettner:2019TL}.
    Right: Confidence intervals obtained from the diffuse astrophysical fit on Asimov data with a power-law spectrum with exponential cutoff.}
    \label{fig:diffuse_spl}
\end{figure}

In order to test and verify our analysis strategy, we calculate the 95\% confidence-level (C.L.) regions for a diffuse neutrino flux based on a single power law 
\begin{equation}
\label{eq:spl}
    \frac{\drv \Phi}{\drv E} = 
    \Phi_0 \cdot \left( \frac{E}{\SI{100}{TeV}} \right) ^ {-\gamma} 
    \frac{\SI{E-18}{}}{\SI{}{\GeV \centi \meter \squared \second \steradian}}.
\end{equation}
We assume the power-law spectrum in Eq.~\eqref{eq:spl} using a normalization of $\Phi_0 = 1.44$ and a spectral index of $\gamma=2.28$~\citep{Stettner:2019TL}, while the background component of atmospheric neutrinos is modeled with \textbf{MCEq}\footnote{\href{https://github.com/afedynitch/MCEq}{https://github.com/afedynitch/MCEq} with hadronic model \textsc{Sibyll-2.3c} \citep{Riehn:2017mfm} and atmosphere \textsc{NRLMSISE-00 Model 2001} \citep{NRLMSISE}.}.
The left panel of Fig.~\ref{fig:diffuse_spl} shows the resulting confidence regions with respect to the best-fit hypothesis assuming the IceCube effective area and a livetime of 10 years.
We find that the overall shape and size of the 95\%~C.L. contour approximately matches the contour presented by IceCube in Ref.~\cite{Stettner:2019TL}.
 As expected, our contour is slightly more stringent, since our analysis does not include systematic uncertainties.
For the \plenum contours in the left panel of Fig.~\ref{fig:diffuse_spl}, we added 10 years of Asimov data from IceCube to 10 years of Asimov data from \plenum-1 or \plenum-2.
Thus, we expect a significant reduction of the uncertainties of the flux parameters, if the systematic uncertainties remain sub-dominant with respect to the statistical uncertainties.

\begin{figure}
    \centering
    \includegraphics[width=0.49\textwidth]{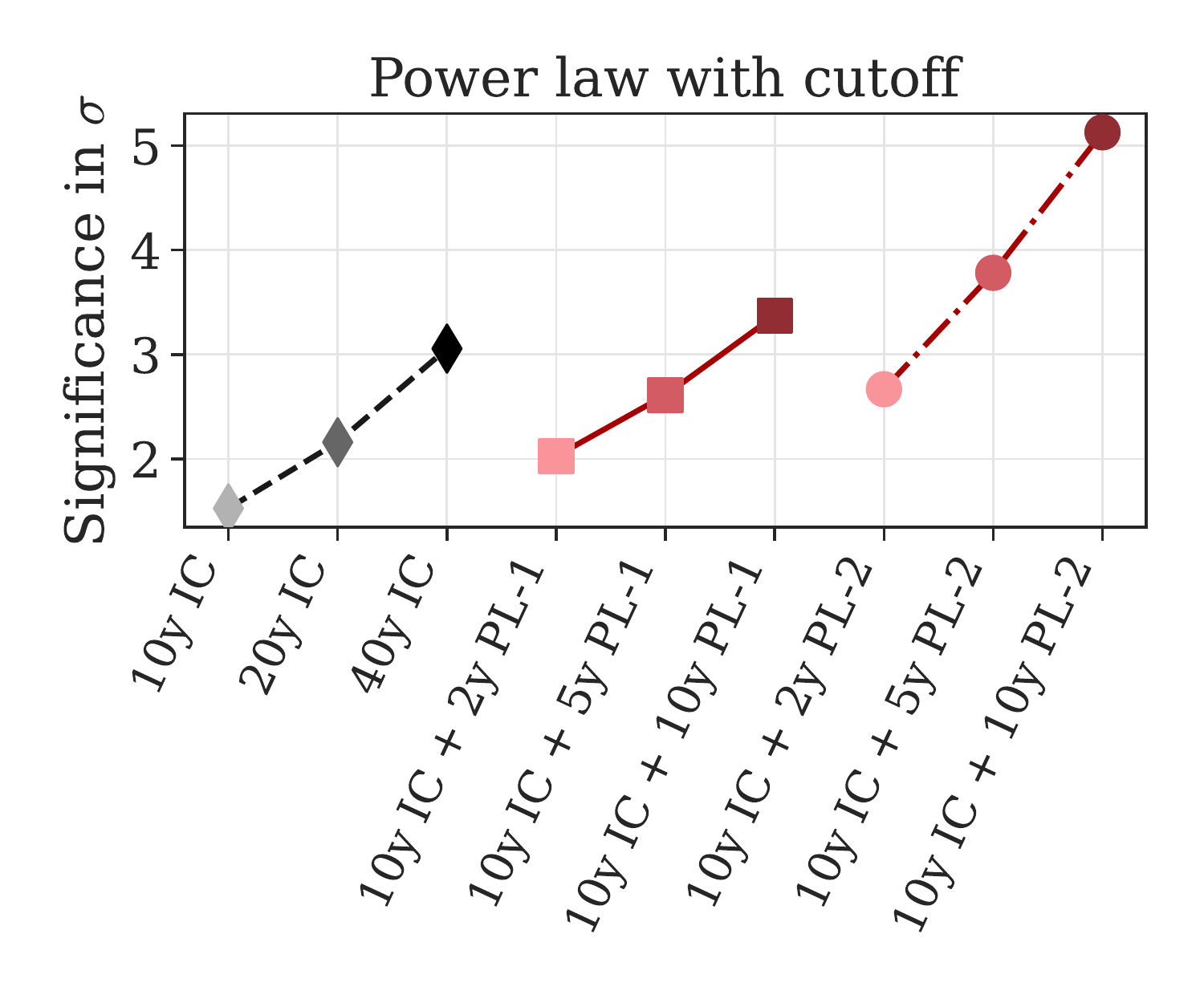}
    \includegraphics[width=0.49\textwidth]{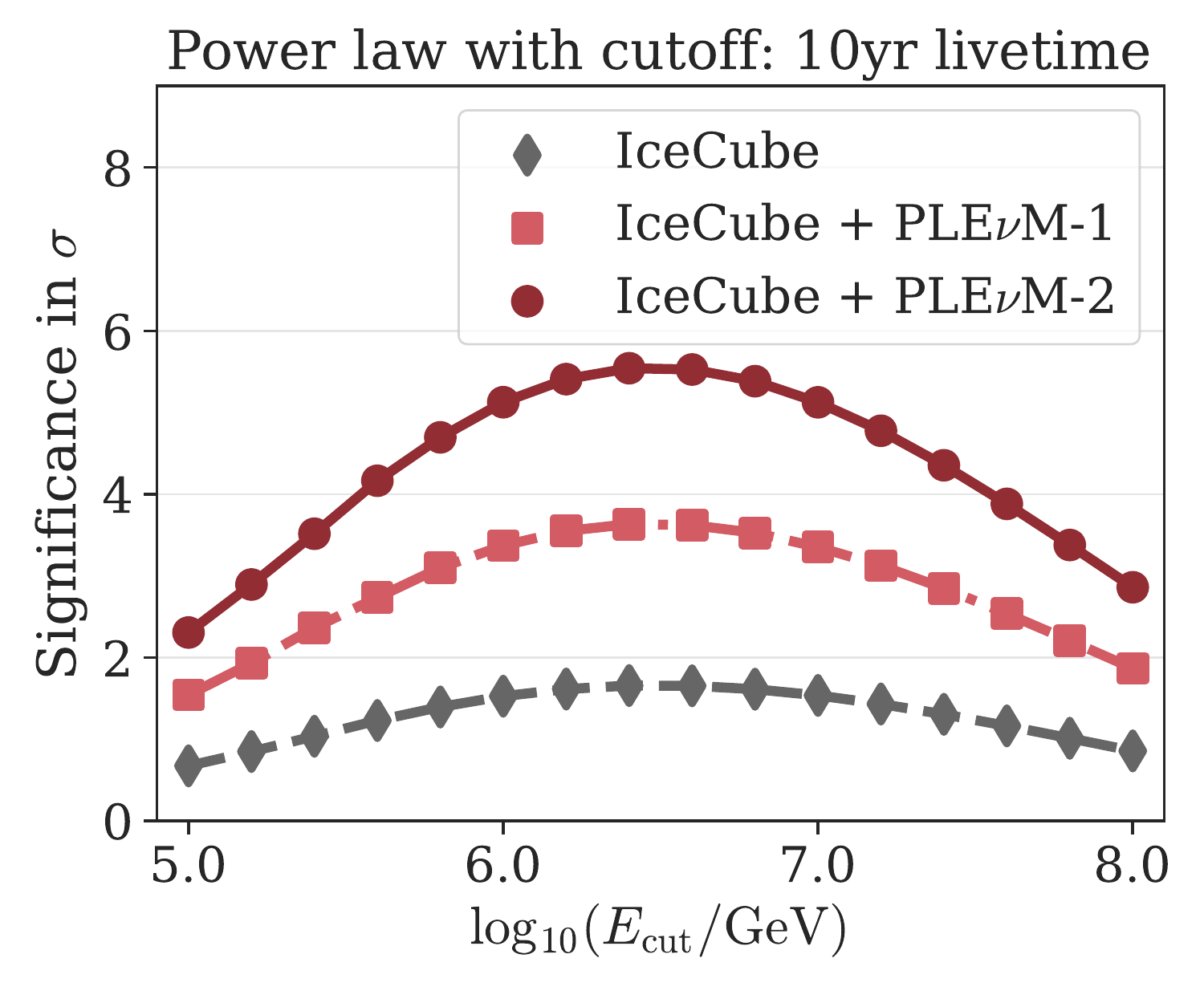}
    \caption{Significance of a likelihood ratio test with a pure power law as null hypothesis and a power law with exponential cutoff as alternative hypothesis.
    Left: Significance as a function of different detector combinations and livetimes.
    Right: Significance as a function of the cutoff energy for different detector combinations and 10 years of livetime each.
    }
    \label{fig:cutoff-significance}
\end{figure}

This leads us to the question of whether \plenum will be able to detect deviations from the pure power-law spectrum.
As a case study, we investigate a power-law spectrum with an exponential cutoff in energy modeled as
\begin{equation}
\label{eq:ecut}
    \frac{\drv \Phi}{\drv E} = 
    \Phi_0 \cdot \left( \frac{E}{\SI{100}{TeV}} \right) ^ {-\gamma} \cdot \exp \left( -\frac{E}{\ecut}\right)
    \frac{\SI{E-18}{}}{\SI{}{\GeV \centi \meter \squared \second \steradian}}.
\end{equation}
As baseline parameters, we choose a cutoff energy of $\ecut=\SI{1}{PeV}$, a spectral index of $\gamma=2.0$ and a flux normalization of $\Phi_0=1.5$.
With IceCube alone, using an Asimov data set with 10 years of livetime based on IceCube's effective area and Eq.~\eqref{eq:ecut}, we find a p-value of 13\% when comparing the cutoff hypothesis with baseline parameters to a pure power-law hypothesis.
Using \plenum-1 and \plenum-2 instead, we find a significance of around $3\sigma$ and $5\sigma$, respectively.
The right panel of Fig.~\ref{fig:diffuse_spl} shows the 95\%~C.L. region for IceCube, \plenum-1, and \plenum-2 for a power-law spectrum with a cutoff.
The left panel of Fig.~\ref{fig:cutoff-significance} shows the evolution of significance with different livetimes and detector configurations.
We conclude that a detector configuration similar to \plenum-2 is necessary in order to claim a deviation from a pure power-law spectrum within 10 years in this exemplary case.
In addition, we calculate the expected significance using \plenum data with cutoff energies between \SI{100}{\TeV} and \SI{10}{\PeV}, while keeping $\gamma=2.0$.
The right plot of Fig.~\ref{fig:cutoff-significance} shows that
the prospects of discovering a cutoff are best for cutoff energies between \SIrange{1}{10}{\PeV}.

\section{Conclusion}
We have presented \plenum, a concept for a global effort to combine and share high-energy neutrino observations with present and future neutrino telescopes.
The field of view of single telescopes is limited, 
but a network of telescopes all over the world will reach a uniform exposure over the whole sky.
The first steps presented here are performance studies based on the assumption of IceCube-like detectors at the locations of IceCube, P-ONE, KM3NeT and Baikal-GVD.
We find that both the number of neutrinos and the discovery potential for point-like neutrino sources
improve by up to three orders of magnitude in the Southern Hemisphere with respect to IceCube's performance,
when three telescopes on the Northern Hemisphere are in operation. 
In order to improve the performance in the Northern Hemisphere, a large detector in the South like IceCube-Gen2 is needed.
Regarding the diffuse flux of astrophysical neutrinos, we investigated the capability of \plenum to distinguish a pure power-law spectrum from a power-law spectrum with a cutoff in energy.
We find that about 10 times the exposure of IceCube is needed to reach a $5\sigma$ rejection of the pure power law, after 10 years of livetime.
This increase of exposure can be reached when IceCube, IceCube-Gen2, P-ONE, KM3NeT and Baikal-GVD are in operation.

Based on the promising results presented in this proceeding, we will continue to refine our methods. 
In future works, we will investigate the advantages of the globally distributed telescopes contributing to \plenum for time-dependent and real-time searches for neutrino sources.
Motivated by the recent discovery of gamma-ray PeVatrons by LHAASO~\cite{caoUltrahighenergyPhotonsPetaelectronvolts2021}, will plan to investigate prospects for identifying Galactic, hadronic sources with \plenum.

\bibliographystyle{JHEP}
\bibliography{references}

\end{document}